# Superconductivity in the Face Centered Cubic W-M-Rh-Ir-Pt M={Mo, Nb, Ta, Re} High Entropy Alloy

Denver Strong and R. J. Cava
*Department of Chemistry Princeton University, Princeton NJ*

## Abstract

We report single phase superconducting face centered cubic (FCC) intermetallic high entropy alloys (HEAs) synthesized via splat cooling. The single phase materials fall at electron counts in the HEA superconductor alloy family where structural stability and optimal superconducting electron counts clash. The materials' superconducting properties follow the general trends published for metallic alloys. Insights are provided as to why an FCC structure may be stable.

## Introduction

High entropy alloys (HEAs) are one of the newest material classes, being studied for fewer than fifteen years.[1,2] They are typically defined as consisting of five or more elements, each of less than 35% concentration, ensuring that no single element dominates the observed properties.[3] Even assuming that element selection is restricted to the 30 transition metals, and that a five-component HEA should be equimolar, then there are on the order of $10^6$ options for such alloys. If including elements beyond the transition metals, and varying their contents from equimolar, the possibilities are nearly endless - a particularly useful characteristic when tuning an alloy to obtain a specific property.

Although in practice the thermodynamics are more nuanced, the basic idea is straightforward. Using equation 1,

$$\Delta G_{mix} = \Delta H_{mix} - T\Delta S_{mix} , \tag{1}$$

the goal for formation of HEA's is to make the term in the above expression that represents the product of temperature times the entropy of mixing ($\Delta S_{mix}$) very large. Including more elements increases the entropy of mixing. Paired with a high temperature T but not limited to those achievable by melting in the laboratory, the second term overcomes any of the enthalpic effects of the first ($\Delta H_{mix}$), and a solid solution can sometimes form.[2,3] When more elements are available, and faster cooling rates from high temperature are employed, (which maximizes the impact of $\Delta S_{mix}$); sometimes there is no segregation into a mixture of phases whose stability is dominated by a negative $\Delta H_{mix}$ and a high entropy solid solution can form on a simple lattice. These solid solutions are typically body-centered cubic (BCC), hexagonal close packed (HCP), or face-centered cubic (FCC), but more recent papers indicate that entropically stabilized solid solutions can form topologically close packed structures as well and are not necessarily straightforward.[4-8]

HEAs can display a wide variety of physical properties, including magnetism, good corrosion resistance, excellent mechanical properties, and superconductivity.[3,9] However, a superconducting face-centered cubic solid solution has not been reported to date.[10] A superconducting FCC HEA is desired for applications due to a high ductility associated with the increased number of slip systems in the structure.[11] FCC superconductors haven't completely eluded researchers, however, because they exist for binary alloys; but they have been difficult to observe within the high entropy regime. Why this has been difficult is revealed in **Figure 1**. The face-centered cubic structure is stable, following a hard sphere model, when more than 8 valence electrons per atom (i.e. the VEC) are available.[12] Below that, a change of the stacking from FCC type to HCP type is expected to occur. This has been explained as resulting from increased hybridization.[12] The stable FCC VEC is not one that is favored for superconductivity, however, and thus the co-existence of HEA superconductivity and the FCC structure has not been observed. This issue is illustrated in **Figure 1**, which shows the difficulties in finding a high entropy superconducting single phase FCC alloy - the FCC structure is only stable for VECs that are outside of the superconducting VEC regime. There are not yet a sufficient number of anomalies known to consistently shift the maximum VEC threshold for superconductivity up in electron count or the VEC threshold for FCC stability down in electron count for both superconductivity and an FCC crystal structure to co-exist in the same material.

Here, we report high entropy alloys derived from the anomalous binary FCC solid solution, $W_{67}Pt_{33}$, which superconducts at 3.0 K.[13] The M-W-Rh-Ir-Pt alloys studied here contain no principle elements, and none of the elements by itself superconducts above 1 K, yet our FCC alloys have a maximum $T_c$=1.6 K. The VEC bounds on structure suggest that an HCP alloy should form rather than an FCC one, and structure-in-structure-out (SISO) methods predict that a BCC structure should form before entering the high entropy regime.[3] Nonetheless we find the FCC structure for these alloys under our synthetic conditions.

This paper is summarized as follows. The next sections describe the synthesis, structural and physical characterization methods. The structural results are described in section 3 while section 4 addresses the superconducting properties. Finally, section 5 summarizes the physical properties, and addresses potential future work.

**Methods**

The samples were made from elemental powders, total weight less than 200 mg, ground in stoichiometric ratios and compressed in a steel dye to form a pellet. The pellets were arc melted 5x on a water-cooled copper hearth with I~50 A and V~480 V, in an argon atmosphere after a zirconium getter was used to remove any residual oxygen from the system. This procedure ensures that the elements are well mixed. Samples were then placed in a copper well with a zirconium getter again. The hearth was then heated with I~50 A for 5-8 seconds before immediately aiming the arc toward the sample in the bottom of the copper well. Heating continued for another 5-8 seconds after the molten state had been reached. The tungsten electrode was then removed from the well and a spring-loaded copper cooling tip immediately plunged into the molten liquid, quenching it into the solid state. The result was thin solidified sheet (t<0.6 mm) samples that could be broken into flakes. The "best" (i.e. single phase) samples were the thinnest after the splat quench.

Powder X Ray diffraction (pXRD) patterns from the HEAs thus obtained were collected in the Bragg Brentano geometry on a Bruker D8 Advance ECO diffractometer, using Cu Kα X=Rays. Resistivity and heat capacity measurements were performed on the largest and smallest remaining flakes, respectively, in a Quantum Design Physical Property Measurement System (PPMS), via a standard 4 probe resistivity puck on an adiabatic refrigeration (ADR) attachment, and a $^3$He heat capacity puck.

**Results**

**Figure 2** shows the pXRD patterns for select samples of $W_{2-x}Mo_xRhIrPt_2$ ($W_{33-x}Mo_xRh_{17}Ir_{17}Pt_{33}$), with the $x$ values selected to result in alloy compositions of 0, 5, and 10% mole fractions of Mo. All the samples display clear Kα1/Kα2 splitting for the diffraction lines, indicative of highly crystalline uniform alloys. In the 10% Mo sample, for example, the nominal concentration is $W_{23}Mo_{10}Rh_{17}Ir_{17}Pt_{33}$, which is clearly within the HEA regime. Each pattern was refined using GSAS-II, with lattice parameters as summarized in **Table 2**. The samples are very clearly FCC. There are occasionally minor HCP peaks present and noted, but they represent such a small concentration that they do not interfere with our interpretation of the data. Refinements of the superconducting FCC phases yield less than 1% error when compared to Vegard's law.

It is common practice to consider intermetallic alloys in terms of the Hume-Rothery rules, where weighted variations are used to identify the robustness of the system. The standard equations employed to calculate the allowed variations in atomic radii and electronegativity are, respectively,[14]

$$\delta = 100\sqrt{\sum_{i=1}^{n} c_i\left(1 - \frac{r_i}{\underline{r}}\right)^2}, \quad (2)$$

$$\Delta\chi = \sqrt{\sum_{i=1}^{n} c_i\left(\chi_i - \underline{\chi}\right)^2}. \quad (3)$$

The underlined values are weighted averages while $c_i$, $r_i$, and $\chi_i$ are the concentration, atomic radii, and Pauling electronegativity of each element. δ and $\Delta\chi$ values less than 10% are generally taken as systems that satisfy the Hume-Rothery rules. The values for our system are summarized in **Table 2**.

Finally, the expected enthalpy of mixing can be approximated by

$$\Delta H_{mix} = \sum_{i=1, j\neq i}^{n} 4\Delta H_{ij} c_i c_j, \quad (4)$$

where $\Delta H_{ij}$ is the mixing enthalpy of binary liquid alloys.[14] In the current analysis, the values for $\Delta H_{ij}$ are taken from the literature [15] and are included in the supplementary material. When the FCC superconductors reported here are compared with other high entropy superconducting solid solutions (**Figure 3**) they are seen to display the smallest δ values. However, the negative mixing enthalpies suggest that the alloys are members of a non-ideal, regular solution. This may be one of the best examples known where a largely negative mixing enthalpy can aid in the formation of a solid solution.[3]

**Superconductivity**

Low temperature resistivity data is shown in **Figure 4** along with points for every alloy reported here alongside reproduced values from the original binary.[13] It has been suggested that linear trends in $T_c$ indicate that the electronic density of states' landscape is relatively unchanged as the electron count is varied.[13] The resistivity curves in **Figure 4a** follow the expected trends based on the original W-Pt binary alloy where less W can drastically reduce the $T_c$. **Figure 4b** demonstrates doping of four different metals into $W_{2.5}RhIrPt_2$ where W is replaced with one of Mo, Nb, Ta, or Re. Notice that the 4d metals slightly reduced the $T_c$ and the 5d transition metals result in the opposite. With only 5% concentration, the set of fifth elements have very little effect on the bulk, provided additional phases do not separate from the host and, in effect, only tune the $E_F$ within a very narrow range of the DOS. Figure 4c confirmed these HEAs do not act significantly different from the generalized empirical rules already discussed.

It's interesting to note that while there's no substantial difference when the fifth element is Mo, Nb, or Ta; the materials with Re have an "unusually high" transition temperature. Albeit still ~1 K, $T_c^{Re}$ is ~20% larger than others with 33.5% W content, yet also appears to trail off slower with VEC. It would be beneficial to complete the study of the Re series, but a more consistent splat cooling apparatus should be used to ensure consistency between samples. An interpretation of the enhanced $T_c$ is because Re is the smallest additional element attempted. Although it's heavy, the real space size may influence the phonon density of states and hence the electron-phonon coupling.

Sufficiently thin samples have a 90% to 10% width of $\Delta T_c \sim 0.2$ K which suggests that the increased disorder present in the splat cooled HEAs does not generally increase the electronic inhomogeneity. This is in contrast to reports where a broadening of the superconducting transition is seen for HEAs[17-26] prepared by conventional arc melting - the slower cooling rates for arc-melted samples give time for atoms to form different local structures. Thus the splat cooling synthesis method employed here may be expected to yield more exact critical temperatures for random homogenous mixtures.

We also study the system with the next highest $T_c$, $W_{3.5}RhIrPt_2$ ($W_{46.7}Rh_{13.3}Ir_{13.3}Pt_{26.7}$). A broad transition occurs at T~1.5 K. With the addition of the Schottky anomaly, it's difficult to pinpoint a transition temperature to determine the superconducting jump. However, it's clear from **Figure 5** that the transition is between 1.3 K < T < 2.0 K. The high temperature side was fit to

$$\frac{C}{T} = (5.75 \cdot 10^{-5})T^4 + (6.47 \cdot 10^{-2})T^2 + 2.92, \quad (5)$$

Where the independent variable is $T^2$. From this, we determine the Sommerfeld constant to be 2.92 (mJ/mol.K$^2$), and we can approximate the Debye temperature as[6]

$$\theta_D = \sqrt[3]{\frac{12\pi^4 N_A k_B}{5*6.47\cdot 10^{-2}}} = 311\ \mathrm{K} \qquad (6)$$

each of which are lower than the weighted averages of the individual elements (gamma = 3.25 mJ/mol.K$^2$, $\Theta_D$ = 366 K).

The equal areas approach suggests a $T_c$~1.5 K, and a $\frac{\Delta C}{\gamma T_c} \sim 1.23$. This latter value is only slightly smaller in comparison to what should be seen theoretically for BCS superconductors (1.43), indicating a bulk, BCS superconductor. The electron-phonon interaction can then be quantified using McMillan's equation[29]

$$\lambda_{e-ph} = \frac{1.04 + \mu^* ln(\theta_D/1.45T_c)}{(1-0.62\mu^*) ln(\theta_D/1.45T_c) - 1.04}. \qquad (8)$$

Assuming $\mu^*$=0.13, equation (8) returns $\lambda_{e\text{-}p}$~0.42, which is only slightly below the average and median values of other alloys.[29]

## Discussion and Conclusion

Recently, the mysteries of metallic bonding in transition metals has attracted some interest.[30] We briefly summarize, and then conceptualize, a proposal in the literature for why these alloys form from a microscopic viewpoint. In close packed systems, each Wyckoff cell has two types of energetic cages — tetrahedral (point P in **Figure 6**) and octahedral (point H in **Figure 6**). Between them is a 'ring critical point' on the Wyckoff surface. This means that the point's electron density changes only along one direction which is along the Wyckoff edge. For the early transition metals, the location of this particular critical point begins near the tetrahedral cage, an electron deficient vertex where three edges come together. As we traverse across the *d* element series, the critical point moves toward the octahedral cage – the vertex where four edges come together. With these two well- defined energetic basins, the stability as a function of electron count can be explained by looking at which vertex collects kinetic energy faster with the addition of more electrons.[30] Elements to the left of Rh collect kinetic energy in the tetrahedral cage faster while the FCC elements to the right, including Rh itself, collect energy faster in the octahedral cage. Considering the high temperature synthesis required for formation, the reported alloys must be in a position such that the tetrahedral cage collects kinetic energy faster than the octahedral cage at low temperatures, but the opposite is true at high temperatures. Ultimately, the sample must be sufficiently mixed so the octahedral electron density of FCC formers can flood the tetrahedral cage of HCP formers at high temperatures. Once *all* the shallow cages are filled, the residual energy collects in the octahedral basin. The importance of this kinetic energy has been discussed previously when considering the formation of a pseudogap in metallic alloys.[31]

Although it's similar, superconductivity is onset at a lower temperature in these alloys when compared to the W-Pt binary. It was previously pointed out the alloying in this binary system is not straightforward, because there's an initial decrease in the lattice parameters when W is added to Pt before the trend reverses to follow Vegard's law.[13] Cooper pairs do not form until after this minimum is reached, and the still uncharacterized electronic structure of the low Pt concentration alloy disappears.

We attempt to provide an explanation for this behavior with a toy model. From Table 1, it's known W is larger than Pt, hence additional W is expected to increase the size of the unit cell. However, W is *slightly* more electronegative and its 5d band sits just below that of Pt. Here, we assume there is significant overlap between the bands. In essence, this becomes an electron sink stealing from the Pt matrix. With more electrons, tungsten's electron shells become half filled, shrinking, acting as Re for the time being. During this time, the $E_F$ is still in platinum's d-DOS. However, once the $E_F$ falls to the top of tungsten's d-DOS, it's energetically favorable for the system to hybridize. After the hybridization, W is an intricate part of the electronic structure and no longer acts as a sink; hence, the lattice parameters grow as expected by Vegard's law.[13] The resulting electronic behavior traces the DOS via the $T_c$ versus W concentration. A similar behavior occurs in W-Pd alloys, but the difference between 4d and 5d makes the situation more difficult to visualize.

It's the help of this alloy, along with the fast cooling rates accessible by splat cooling, that allow high quality FCC superconducting HEAs to form. Provided the W-Pt binary is the active agent in superconductivity, it is worthwhile to study the connection between hybridization and superconductivity in other binary transition metal alloys. Further research would include studying how dilute the active binary may be within the HEA regime.

In terms of superconductivity, when compared to previously published trends, these materials stand out due the limited number of superconducting FCC examples. They continue to follow previously reported Tc vs VEC trends, although they do have very narrow resistive transitions when compared to typical HEA superconductors,[17-26] likely due to the synthesis method employed. Superconducting FCC HEAs are interesting systems to study moving forward due to their location between structural and superconducting stability boundaries, work which may help to illuminate the limits of BCS theory.

**Acknowledgement**

This research was funded by the Gordon and Betty Moore Foundation, grant number GBMF-9066

## References

1) Cantor, Brian, et al. "Microstructural development in equiatomic multicomponent alloys." *Materials Science and Engineering: A* 375 (2004): 213-218.
2) Yeh, J-W., et al. "Nanostructured high-entropy alloys with multiple principal elements: novel alloy design concepts and outcomes." *Advanced Engineering Materials* 6.5 (2004): 299-303.
3) Miracle, Daniel B., and Oleg N. Senkov. "A critical review of high entropy alloys and related concepts." *Acta Materialia* 122 (2017): 448-511.
4) Liu, Bin, et al. "Formation and superconductivity of single-phase high-entropy alloys with a tetragonal structure." *ACS Applied Electronic Materials* 2.4 (2020): 1130-1137.
5) Liu, Bin, et al. "Superconductivity and paramagnetism in Cr-containing tetragonal high-entropy alloys." *Journal of Alloys and Compounds* 869 (2021): 159293.
6) Stolze, Karoline, et al. "High-entropy alloy superconductors on an α-Mn lattice." *Journal of Materials Chemistry C* 6.39 (2018): 10441-10449.
7) Liu B, Wu JF, Cui YW, Zhu QQ, Xiao GR, Wang HD, et al. Structural evolution and superconductivity tuned by valence electron concentration in the Nb–Mo–Re–Ru–Rh high-entropy alloys. *J Mater Sci Technol* (2021) 85:11. doi:10.1016/j.jmst.2021.02.002
8) Browne, A.J.; Strong, D.P.; Cava, R.J. Stability and superconductivity of 4d and 5d transitional metal high-entropy alloys. arXiv 2021, arXiv:2109.10043. Available online: https://arxiv.org/abs/2109.10043 (accessed 23 Dec 2022)
9) Lee, Yea-Shine, and Robert J. Cava. "Superconductivity in high and medium entropy alloys based on MoReRu." *Physica C: Superconductivity and its Applications* 566 (2019): 1353520.
10) Ishizu, N.; Kitagawa, J. Trial of a search for a face-centered-cubic high-entropy alloy superconductor. arXiv 2020, arXiv:2007.00788. Available online: https://arxiv.org/abs/2007.00788 (accessed on 23 Dec 2022).
11) Callister Jr., W.D.; Materials Science and Engineering an Introduction. In *Dislocations and Strengthening Mechanisms*; John Wiley & Sons Inc. 2000; pp 153-184.
12) Pettifor, D. G. "Theory of the crystal structures of transition metals." *Journal of Physics C: Solid State Physics* 3.2 (1970): 367.
13) LUO H-L 19683. Less Common Metals 15 299-302

14) Sheng, G. U. O., and Chain Tsuan Liu. "Phase stability in high entropy alloys: Formation of solid-solution phase or amorphous phase." *Progress in Natural Science: Materials International* 21.6 (2011): 433-446.
15) Boer, F.R. de, Mattens, W C.M.,Boom, R, Miedema, A R, and Niessen, A K. Cohesion in metals. Transition metal alloys. Netherlands: N. p., 1988. Web.
16) Troparevsky, M.C, Morris, J.R., Kent, P.R.C, Lupini, A.R., Stocks, G.M., "Criteria for Predicting the Formation of Single-Phase High-Entropy Alloys." *Phys. Rev. X* **5**, 011041 (2015)
17) P. Koželj, S. Vrtnik, A. Jelen, S. Jazbec, Z. Jaglici ˇ c, S. Maiti, ´ M. Feuerbacher, W. Steurer, and J. Dolinšek, Phys. Rev. Lett. 113, 107001 (2014).
18) F. von Rohr, M. J. Winiarski, J. Tao, T. Klimczuk, and R. J. Cava, Proc. Nat. Acad. Sci. USA 113, E7144 (2016).
19) F. O. von Rohr and R. J. Cava, Phys. Rev. Mater. 2, 034801 (2018).
20) Kim, G.; Lee, M.-H.; Yun, J.H.; Rawat, P.; Jung, S.-G.; Choi, W.; You, T.-S.; Kim, S.J.; Rhyee, J.-S. Strongly correlated and strongly coupled s-wave superconductivity of the high entropy alloy Ta1/6Nb2/6Hf1/6Zr1/6Ti1/6 compound. *Acta Mater.* **2020**, *186*, 250–256.
21) Wu, K.-Y.; Chen, S.-K.; Wu, J.-M. Superconducting in equal molar NbTaTiZr-based high-entropy alloys. *Nat. Sci.* **2018**, *10*, 110–124.
22) Marik, S.; Varghese, M.; Sajilesh, K.P.; Singh, D.; Singh, R.P. Superconductivity in equimolar Nb-Re-Hf-Zr-Ti high entropy alloy. *J. Alloys Compd.* **2018**, *769*, 1059–1063.
23) Ishizu, N.; Kitagawa, J. New high-entropy alloy superconductor Hf21Nb25Ti15V15Zr24. *Res. Phys.* **2019**, *13*, 102275.
24) Lee, Y.-S.; Cava, R.J. Superconductivity in high and medium entropy alloys based on MoReRu. *Phys. C* **2019**, *566*, 1353520.
25) Marik, S.; Motla, K.; Varghese, M.; Sajilesh, K.P.; Singh, D.; Breard, Y.; Boullay, P.; Singh, R.P. Superconductivity in a new hexagonal high-entropy alloy. *Phys. Rev. Mater.* **2019**, *3*, 060602(R).
26) Liu, B.; Wu, J.; Cui, Y.; Zhu, Q.; Xiao, G.; Wu, S.; Cao, G.; Ren, Z. Superconductivity in hexagonal Nb-Mo-Ru-Rh-Pd high-entropy alloys. *Scr. Mater.* **2020**, *182*, 109–113.
27) Matthias, Bernd T. "Empirical relation between superconductivity and the number of valence electrons per atom." *Physical review* 97.1 (1955): 74.
28) Toby, B. H., & Von Dreele, R. B. (2013). "GSAS-II: the genesis of a modern open-source all purpose crystallography software package". *Journal of Applied Crystallography*, **46**(2), 544-549.
29) W. L. McMillan, Phys. Rev. 167, 331 1968.
30) S. K. Riddle, T. R. Wilson, M. Rajivmoorthy and M. E. Eberhart, Molecules, 2021, 26, 5396.
31) Mizutani, U.; Sato, H. The physics of the Hume-Rothery electron concentration rule. Crystals 2017, 7, 9.

## List of Tables

**Table 1: Breakdown of elemental characteristics.** [a]Radii, Pauling electronegativity, $T_c$, $\gamma$, and $\theta_D$(0 K) data were taken from https://www.knowledgedoor.com/. [b]The e/a value is related to the calculated number of itinerant electrons obtained via DFT calculations.[31]

| | Radius [pm]* | Pauling Electronegativity[a] | VEC | e/a[b] | $T_c$ [K][a] | $\gamma$ [mJ/(mol.K$^2$)][a] | $\theta_D$(0 K) [K][a] |
|---|---|---|---|---|---|---|---|
| W | 141 | 2.36 | 6 | 1.43 | 0.015 | 1.01 | 383 |
| Rh | 134 | 2.28 | 9 | 1.00 | 0.0325 | 4.65 | 512 |
| Ir | 136 | 2.20 | 9 | 1.60 | 0.014 | 3.14 | 420 |
| Pt | 139 | 2.28 | 10 | 1.63 | < 0.03 | 6.54 | 237 |
| Mo | 140 | 2.16 | 6 | 1.39 | 0.915 | 1.83 | 423 |
| Nb | 147 | 1.6 | 5 | 1.32 | 9.288 | 7.80 | 276 |
| Ta | 147 | 1.5 | 5 | 1.57 | 4.483 | 5.87 | 246 |
| Re | 137 | 1.9 | 7 | 1.40 | 1.697 | 2.29 | 416 |

**Table 2:** A list of all of the newly reported alloys. Shaded yellow alloys are within the high entropy regime. Calculations included here are described in the main text, typically simply weighted standard deviations. $a$ is the lattice parameter for the cubic lattice. $\delta$ describes the variation in atomic radii calculated in equation 2, EN is the weighted electronegativity, $\Delta\chi$ is the weighted standard deviation of the electronegativities, each of which are unitless. $\Delta H_{mix}$ is the mixing enthalpy calculated with equation 4 with source values from reference [15] and $\Delta S_{mix}$ is the change in the mixing entropy for a solid solution. VEC is the valence electron count and is defined as the weighted average of valence electrons, VEC spread is the weighted standard deviation of the former. e/a is the weighted average of itinerant electrons per atom. $T_c$ is the measured critical temperature during resistivity experiments. $W_{2.3}Mo_{0.7}RhIrPt_2$ was not characterized because it had a substantial hcp phase.

| Logical Formula | True Formula | a[A] | δ | EN | Δχ | ΔHmix [kJ/mol] | ΔSmix [J/mol.K] | VEC | VEC spread | e/a[b] | Tc [K] |
|---|---|---|---|---|---|---|---|---|---|---|---|
| W3.5RhIrPt2 | W46.7Rh13.3Ir13.3Pt26.7 | 3.918(1) | 0.99 | 2.31 | 0.02 | -16.40 | 10.36 | 7.87 | 0.23 | 1.45 | 1.56 |
| W3RhIrPt2 | W42.9Rh14.3Ir14.3Pt28.6 | 3.910(1) | 1.02 | 2.30 | 0.02 | -16.17 | 10.62 | 8.00 | 0.22 | 1.45 | 1.10 |
| W2.65Mo0.35RhIrPt2 | W37.9Mo5Rh14.3Ir14.3Pt28.6 | 3.911(1) | 1.03 | 2.29 | 0.03 | -16.94 | 11.90 | 8.00 | 0.22 | 1.45 | 1.25 |
| W2.3Mo0.7RhIrPt2 | W32.9Mo10Rh14.3Ir14.3Pt28.6 | * | 1.03 | 2.28 | 0.03 | -17.72 | 12.55 | 8.00 | 0.22 | 1.45 | * |
| W2.5RhIrPt2 | W38.5Rh15.4Ir15.4Pt30.8 | 3.904(1) | 1.06 | 2.30 | 0.02 | -15.68 | 10.86 | 8.15 | 0.21 | 1.45 | 0.90 |
| W2.175Mo0.325RhIrPt2 | W33.5Mo5Rh15.4Ir15.4Pt30.8 | 3.904(1) | 1.07 | 2.29 | 0.03 | -16.51 | 12.09 | 8.15 | 0.21 | 1.45 | 0.85 |
| W1.85Mo0.65RhIrPt2 | W28.5Mo10Rh15.4Ir15.4Pt30.8 | 3.901(1) | 1.07 | 2.28 | 0.03 | -17.34 | 12.69 | 8.15 | 0.21 | 1.45 | 0.85 |
| W2RhIrPt2 | W33.3Rh16.7Ir16.7Pt33.3 | 3.898(1) | 1.11 | 2.29 | 0.02 | -14.77 | 11.06 | 8.33 | 0.20 | 1.45 | 0.45 |
| W1.7Mo0.3RhIrPt2 | W28.3Mo5Rh16.7Ir16.7Pt33.3 | 3.896(1) | 1.11 | 2.28 | 0.03 | -15.67 | 12.23 | 8.33 | 0.20 | 1.45 | 0.40 |
| W1.4Mo0.6RhIrPt2 | W23.3Mo10Rh16.7Ir16.7Pt33.3 | 3.894(1) | 1.11 | 2.27 | 0.03 | -16.57 | 12.75 | 8.33 | 0.20 | 1.45 | 0.30 |
| W2.175Ta0.325RhIrPt2 | W33.5Ta5Rh15.4Ir15.4Pt30.8 | 3.912(1) | 1.45 | 2.26 | 0.08 | -21.19 | 12.1 | 8.11 | 0.22 | 1.46 | 0.90 |
| W1.7Ta0.3RhIrPt2 | W28.3Ta5Rh16.7Ir16.7Pt33.3 | 3.905(1) | 1.48 | 2.25 | 0.08 | -20.63 | 12.23 | 8.28 | 0.21 | 1.46 | 0.40 |
| W2.175Nb0.325RhIrPt2 | W33.5Nb5Rh15.4Ir15.4Pt30.8 | 3.911(1) | 1.44 | 2.26 | 0.07 | -21.38 | 12.1 | 8.11 | 0.22 | 1.45 | 0.80 |
| W1.7Nb0.3RhIrPt2 | W28.3Nb5Rh16.7Ir16.7Pt33.3 | 3.907(1) | 1.46 | 2.25 | 0.07 | -20.82 | 12.23 | 8.28 | 0.21 | 1.45 | 0.40 |
| W2.175Re0.325RhIrPt2 | W33.5Re5Rh15.4Ir15.4Pt30.8 | 3.900(1) | 1.07 | 2.28 | 0.04 | -14.26 | 12.1 | 8.21 | 0.21 | 1.45 | 1.00 |
| W1.85Re0.65RhIrPt2 | W28.5Re10Rh15.4Ir15.4Pt30.8 | 3.895(1) | 1.08 | 2.25 | 0.06 | -12.75 | 12.69 | 8.26 | 0.2 | 1.45 | 1.05 |
| W1.7Re0.3RhIrPt2 | W28.3Re5Rh16.7Ir16.7Pt33.3 | 3.896(1) | 1.11 | 2.27 | 0.04 | -13.16 | 12.23 | 8.38 | 0.2 | 1.45 | 0.65 |

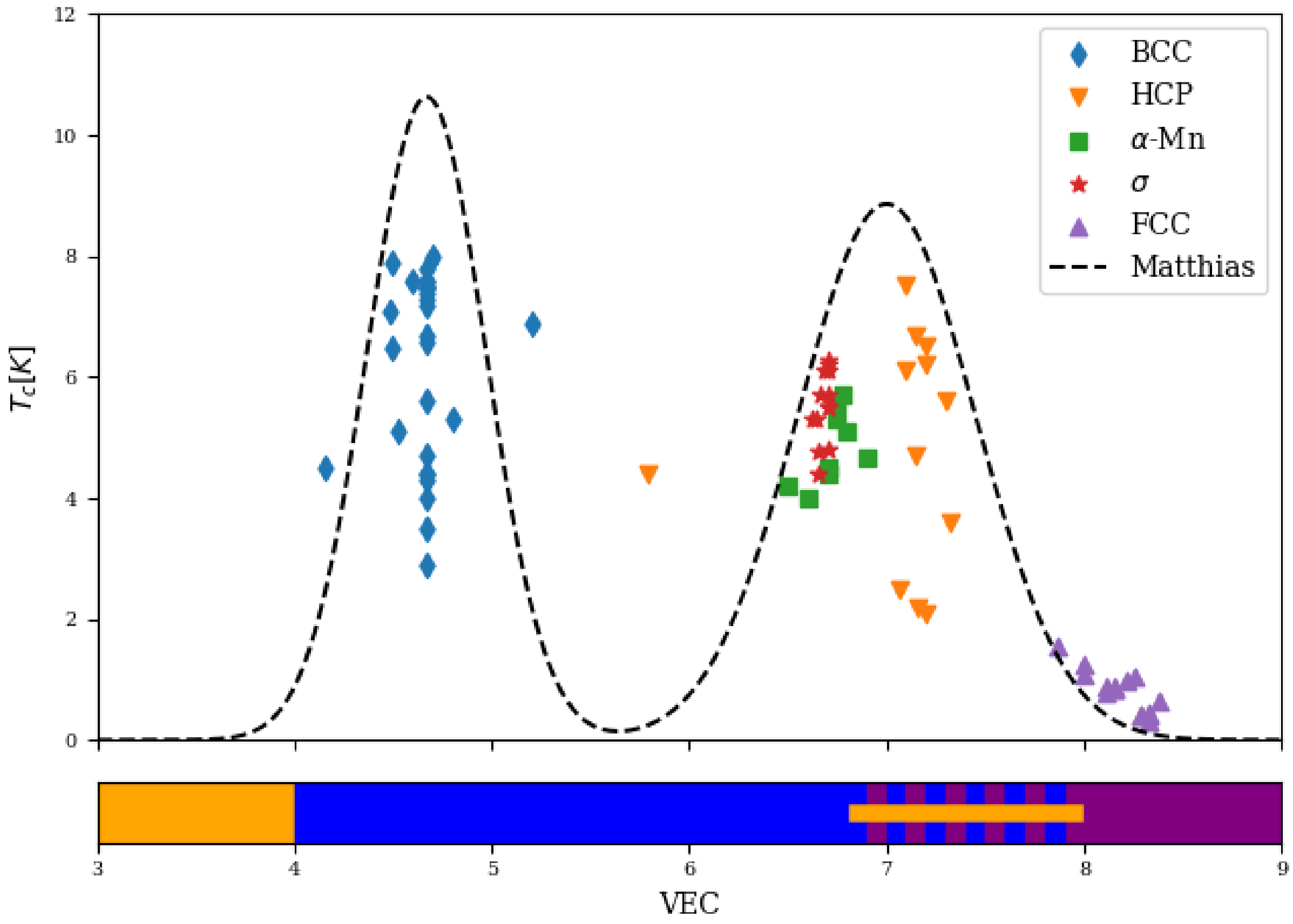

**Figure 1: Overlaid plots of Tc and structure stability vs VEC.** The scatter points are a collection of previously reported critical temperatures for high entropy alloys (HEA) along with the contents from this work.[17-26] The dashed line shows the empirical 'double hump' trend for crystalline metal alloys versus valence electron count (VEC).[27] Calculations for single elements show that the BCC crystal structure is the most stable between VECs of 4 to 6.9 (blue rectangle).[12] The purple rectangle displays the VEC range where FCC structures are stable, while the yellow is HCP. The region below is an accumulation of experimental stability data. The overlay of the two plots suggests there are not enough anomalies to make a FCC superconductor.

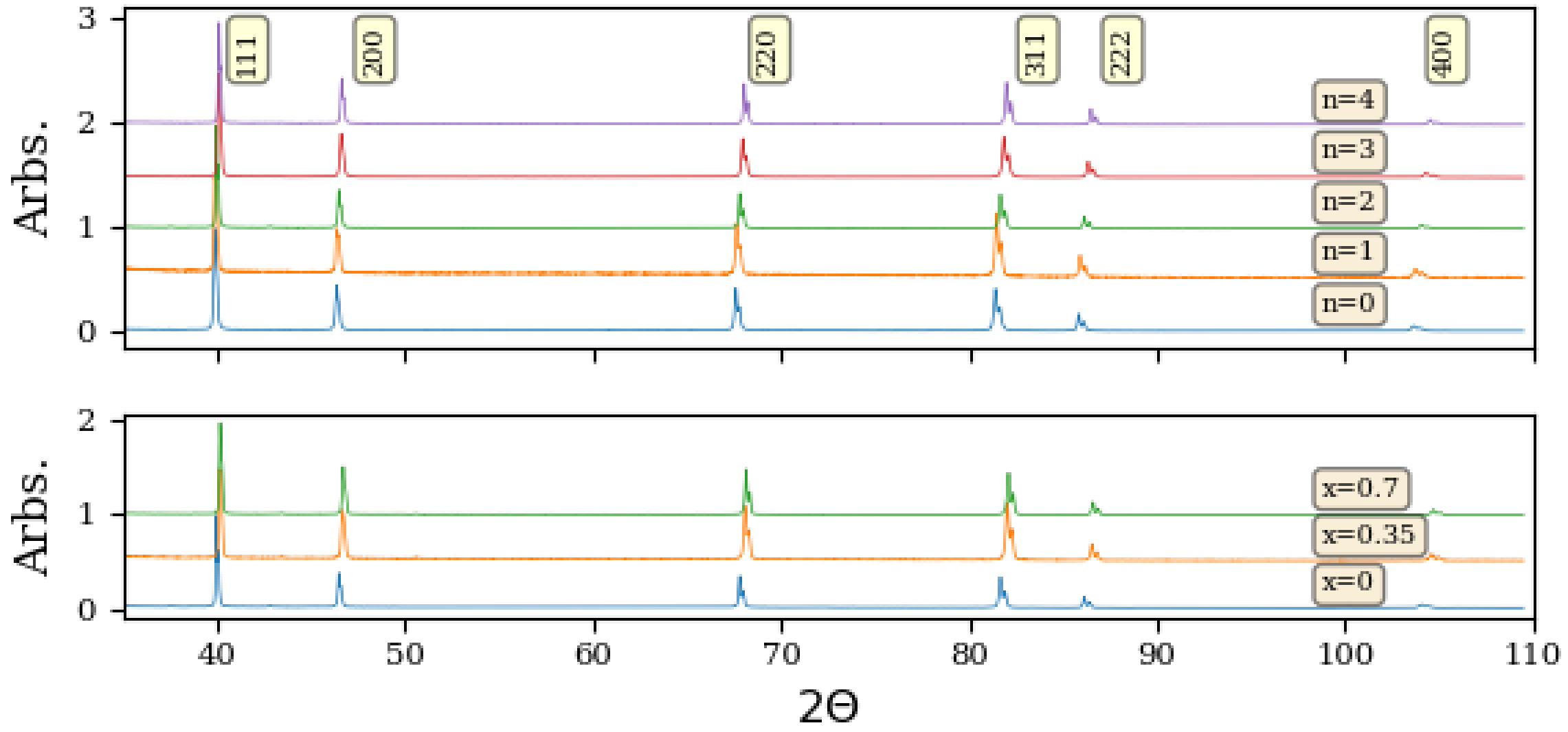

**Figure 2: Diffractograms (Cu Kα radiation) at ambient temperature for splat cooled $W_{4-n/2}RhIrPt_2$ and $W_{3-x}Mo_xRhIrPt_2$.** The results show clear $k\alpha_1/k\alpha_2$ splitting in an FCC lattice. A Le bail refinement done with GSAS-II finds the lattice parameters summarized in Table 1.[28] Similar data for $W_{2.5-x}Mo_xRhIrPt_2$ and $W_{2-x}Mo_xRhIrPt_2$ are included in the supplemental information.

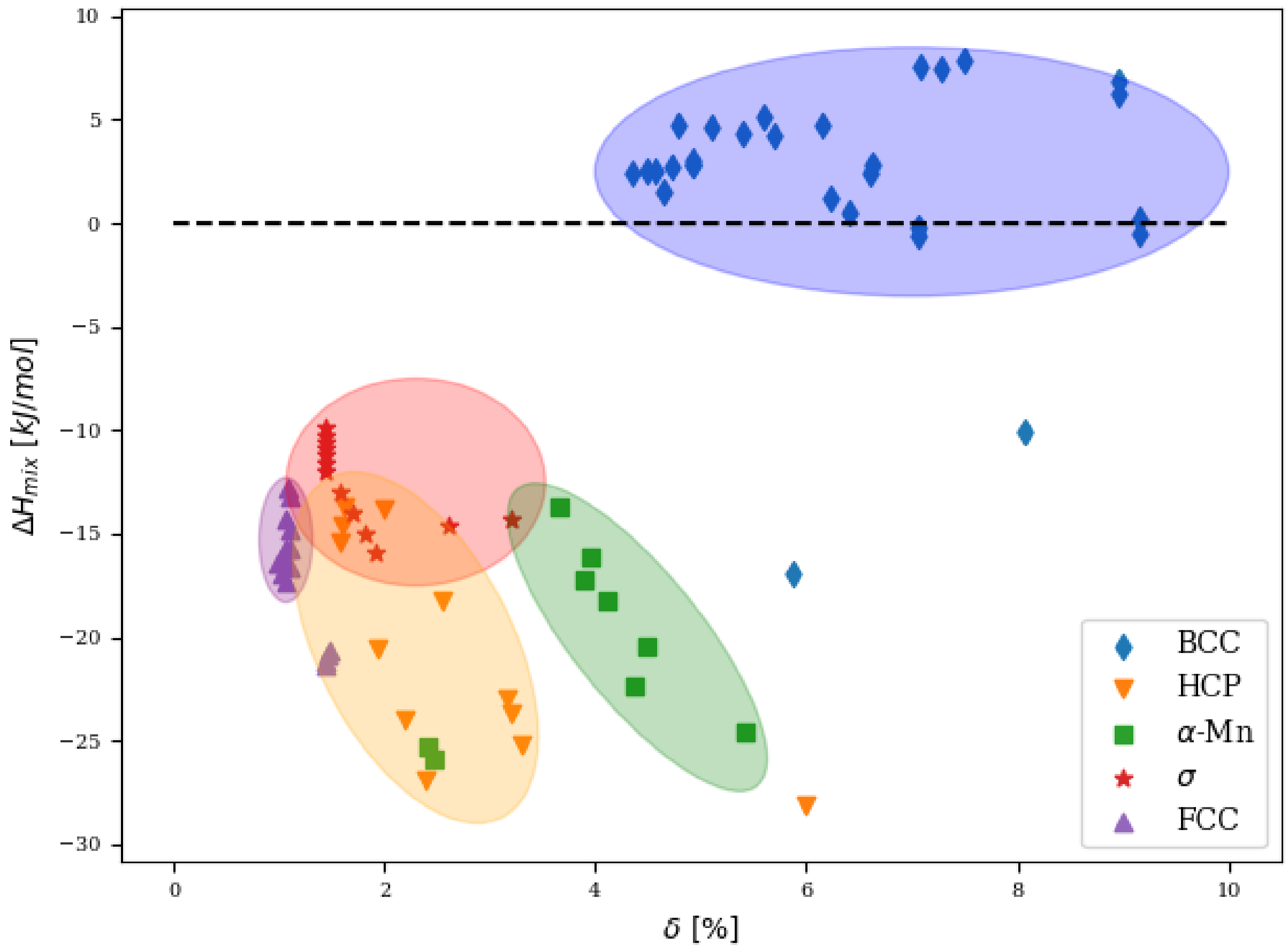

**Figure 3:** High entropy superconducting solid solutions. Plot inspired by reference [3] but limited to solid solution superconductors from refs [17-26] and this work. The FCC alloys are reported in this paper. Solid solutions typically form with small deviations in the atomic radii ($\delta$[%] - defined in equation 2) and small mixing enthalpies ($\Delta H_{mix}$ - defined in equation 4). The original Hume-Rothery rules set a boundary below 15% deviation in atomic radii,[3] but here suggest to stay below 10%. The updated boundaries for $\Delta H_{mix}$ are wider than the original plot along with other reports attempting to keep the value near 0 kJ/mol.[3,16] It's interesting to note that BCC superconducting alloys seem to be separate from the rest from other superconducting materials (where more negative mixing enthalpies are found regardless of atomic size).

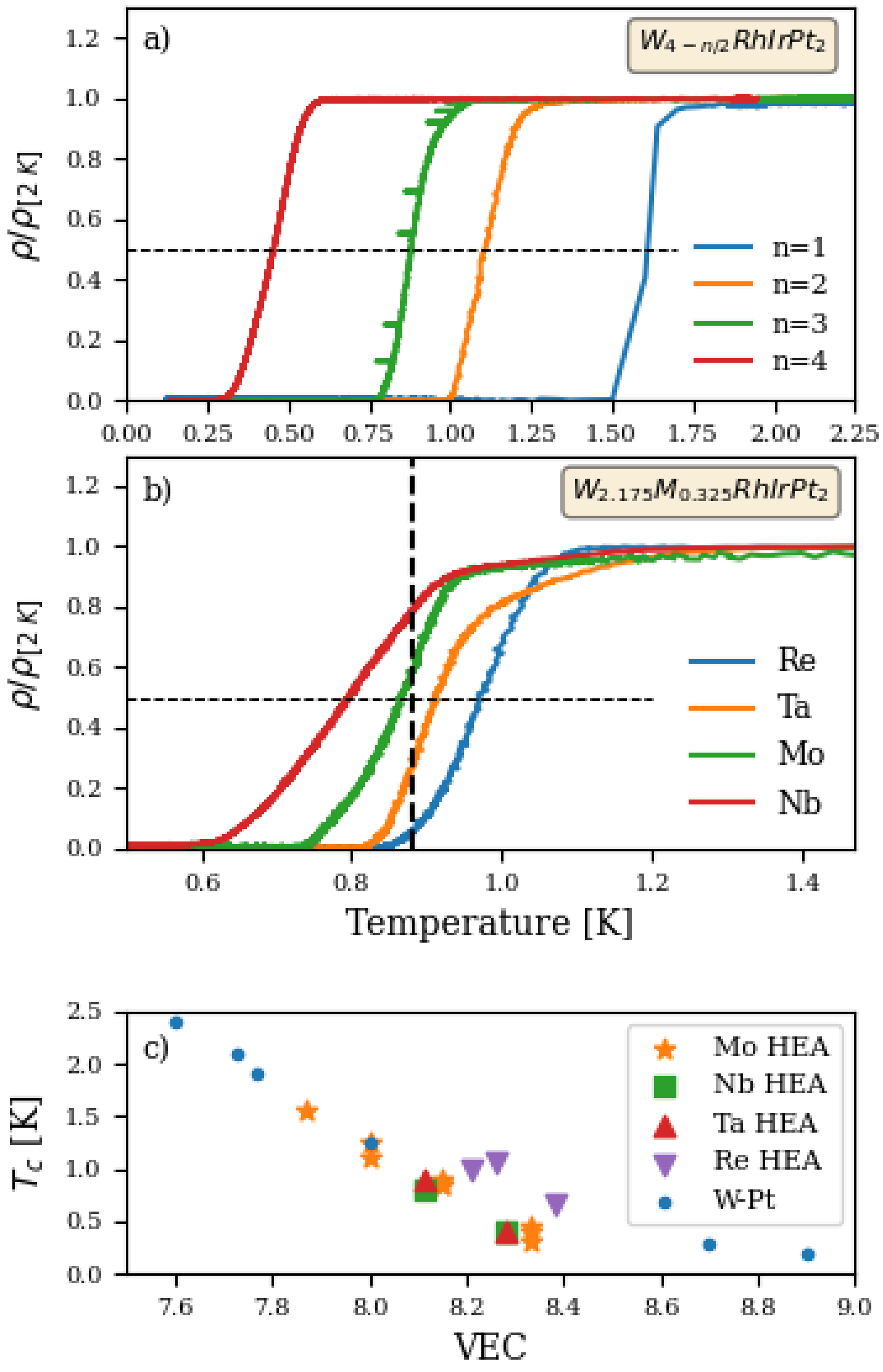

**Figure 4: Brief Summary of Newest HEA alloys. a)** Resistivity curves for pure phase $W_{4-n}RhIrPt_2$ superconductors. Less W clearly decreases the transition temperature as expected. **b)** Resistivity curves for $W_{2.175}M_{0.325}RhIrPt_2$ M={Re, Ta, Mo, Nb}. The additional elements have similar resistivity curves and transition temperatures which deviate slightly from $W_{2.5}RhIrPt_2$ indicated by the vertical dashed line. **c)** All newly reported HEA $T_c$s in comparison with the original W-Pt superconductors versus valence electron count (VEC). The new alloys follow known trends. Re samples warrant further study due to their enhanced $T_c$.

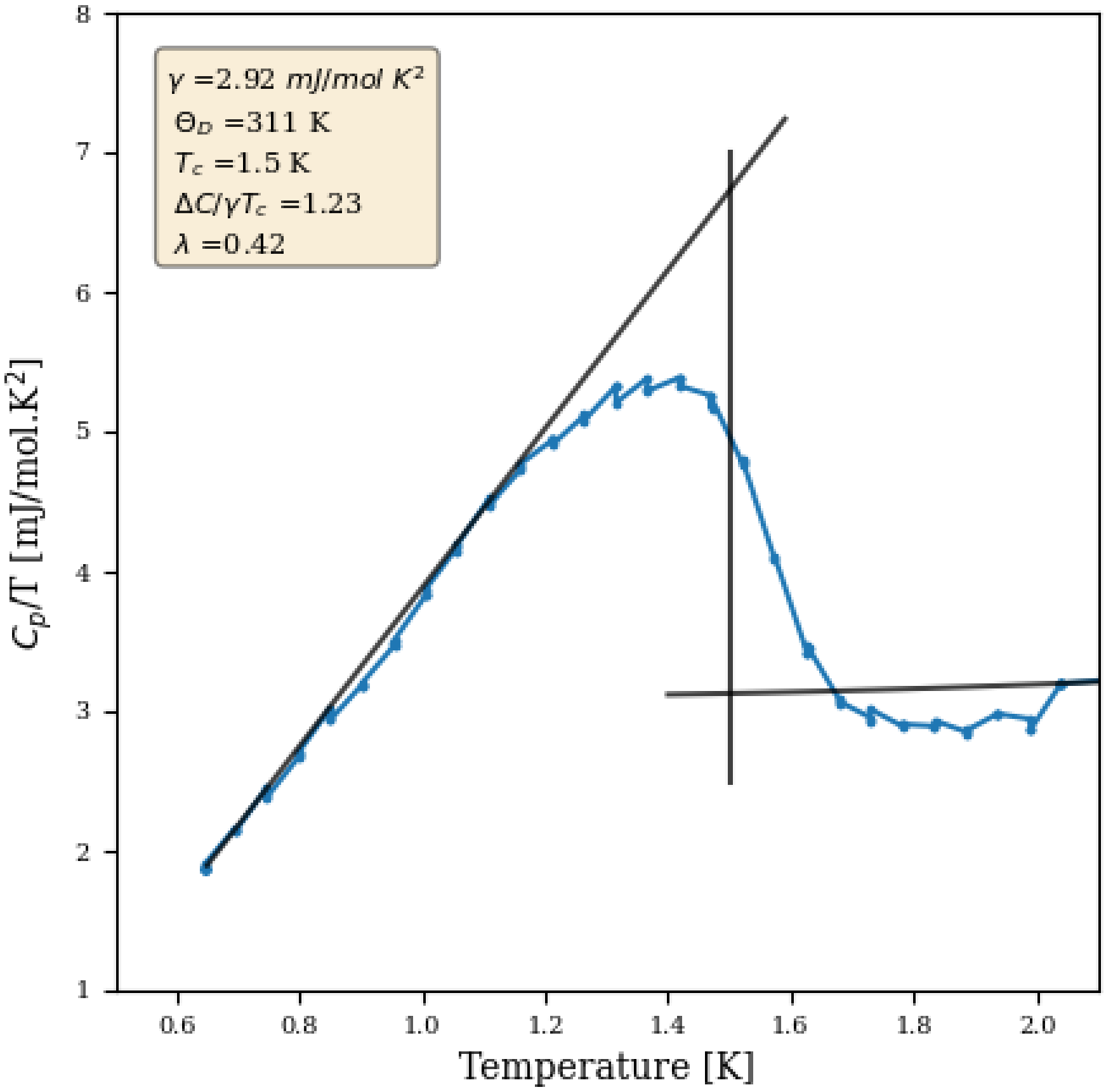

**Figure 5: [3]He data and analysis.** Heat capacity DATA AND ANALYSIS for the W-rich alloy, $W_{3.5}RhIrPt_2$. There's a clear, broad superconducting transition following an anomaly. The high temperature region was fit to equation (5) and the low temperature to equation (7) resulting in $\Delta C_p/\gamma T_c \sim 1.23$, which is slightly low, but is paired with a low electron-phonon coupling of 0.42 calculated via equation (8).

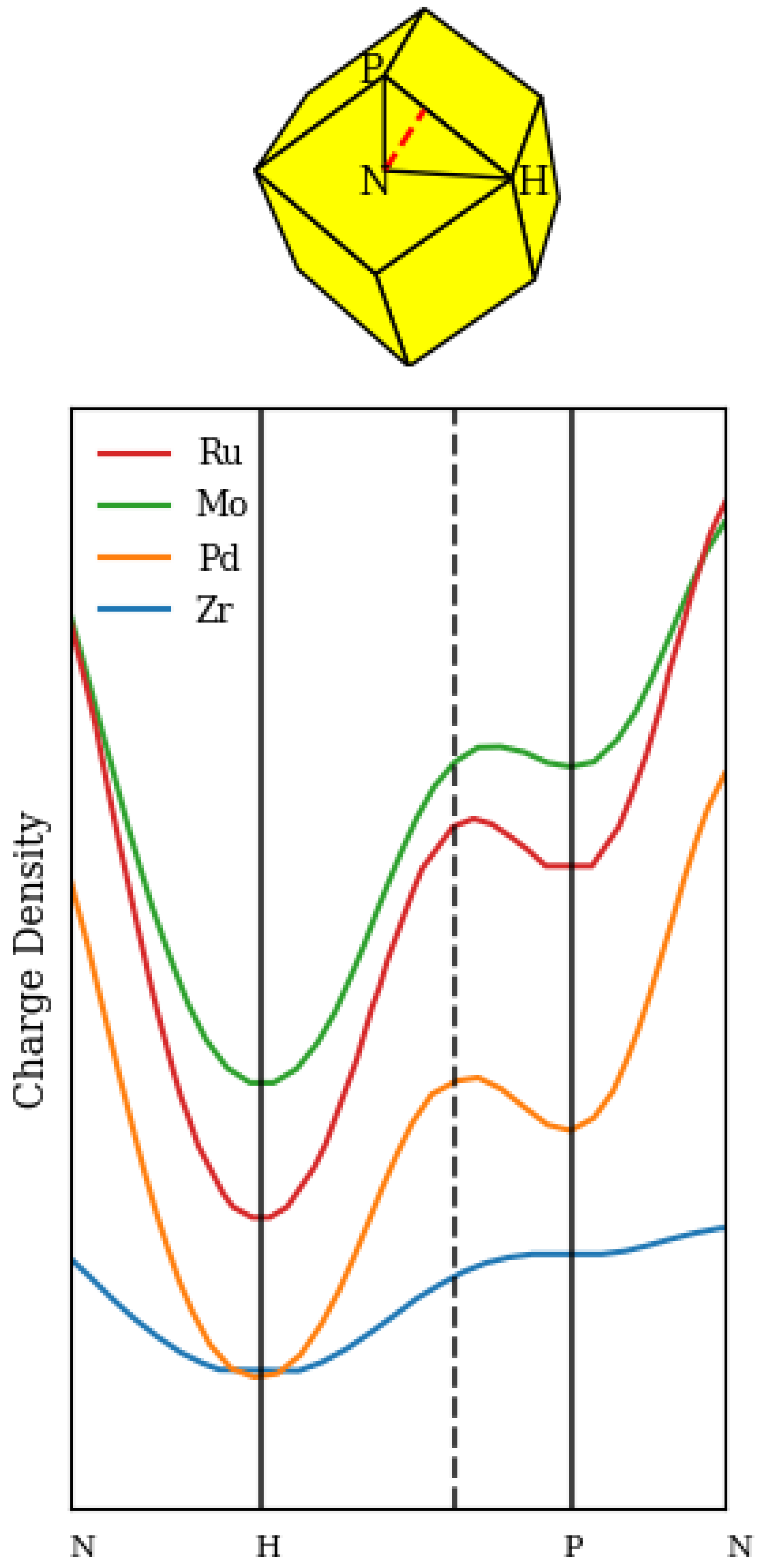

**Figure 6:** Calculated charge density at high symmetry points in the brillouin zone for different 4*d* metals constrained to an FCC structure, reproduced from ref [30]. High symmetry point P denotes a tetrahedral cage which contains electrons. Likewise, an octahedral cage exists at point H. A ring critical point exists ⅓ the way from P to H and is shown with a dashed line. This plot demonstrates an electron's desire to fill the tetrahedral cage, point P, first, before spilling into the octahedral point H which stabilizes the FCC structure.

**Supplemental:**

*Enthalpy of Mixing values used in calculations (ref Cohesion in metals. Transition metal alloys by F.R. de Boer)*

| kJ/mol | W | Rh | Ir | Pt | Mo | Nb | Ta | Re |
|---|---|---|---|---|---|---|---|---|
| W | 0 | -9 | -16 | -20 | 0 | -8 | -7 | -4 |
| Rh | -9 | 0 | 1 | -2 | -15 | -46 | -45 | 1 |
| Ir | -16 | 1 | 0 | 0 | -21 | -53 | -52 | -3 |
| Pt | -20 | -2 | 0 | 0 | -28 | -67 | -66 | -4 |
| Mo | 0 | -15 | -21 | -28 | 0 | - | - | - |
| Nb | -8 | -46 | -53 | -67 | - | 0 | - | - |
| Ta | -7 | -45 | -52 | -66 | - | - | 0 | - |
| Re | -4 | 1 | -3 | -4 | - | - | - | 0 |

*Additional Diffractograms*

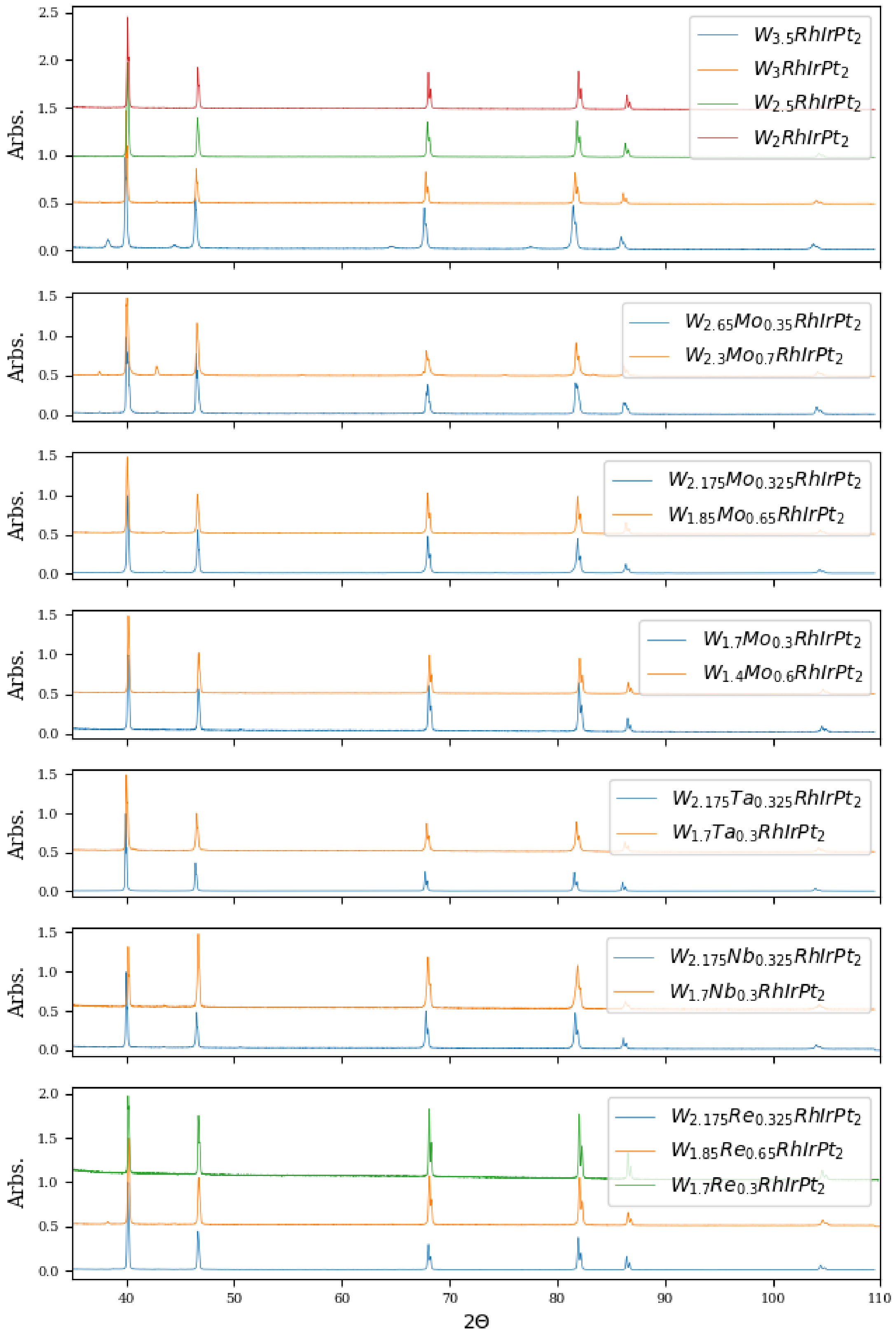

*ADR Data*

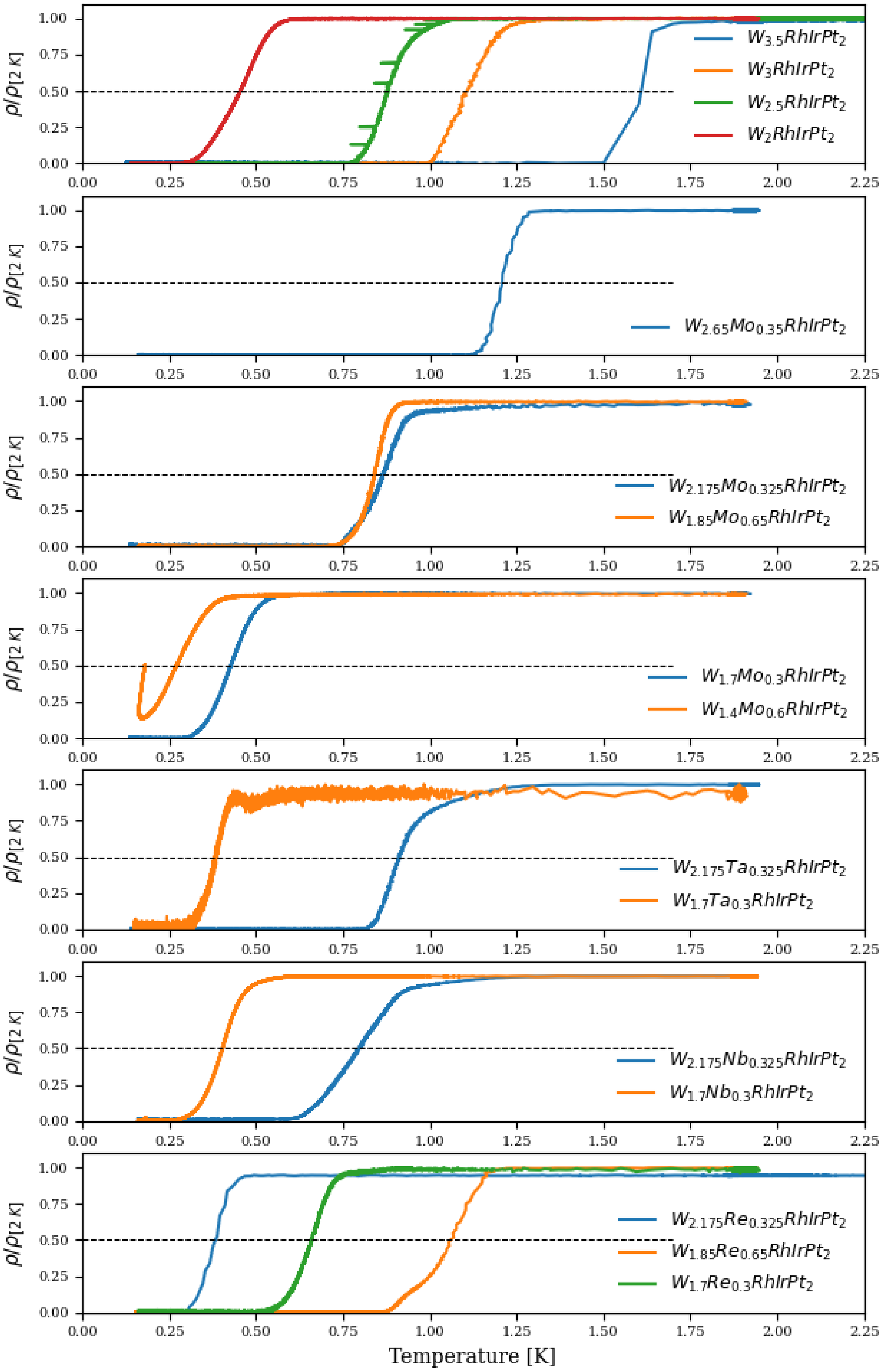

$\rho/\rho_{[2\ K]}$
$W_{3.5}RhIrPt_2$
$W_3RhIrPt_2$
$W_{2.5}RhIrPt_2$
$W_2RhIrPt_2$
$W_{2.65}Mo_{0.35}RhIrPt_2$
$W_{2.175}Mo_{0.325}RhIrPt_2$
$W_{1.85}Mo_{0.65}RhIrPt_2$
$W_{1.7}Mo_{0.3}RhIrPt_2$
$W_{1.4}Mo_{0.6}RhIrPt_2$
$W_{2.175}Ta_{0.325}RhIrPt_2$
$W_{1.7}Ta_{0.3}RhIrPt_2$
$W_{2.175}Nb_{0.325}RhIrPt_2$
$W_{1.7}Nb_{0.3}RhIrPt_2$
$W_{2.175}Re_{0.325}RhIrPt_2$
$W_{1.85}Re_{0.65}RhIrPt_2$
$W_{1.7}Re_{0.3}RhIrPt_2$
0.00 0.25 0.50 0.75 1.00
0.00 0.25 0.50 0.75 1.00 1.25 1.50 1.75 2.00 2.25
Temperature [K]

*ADR explanation*

The tail for $W_{1.4}Mo_{0.6}RhIrPt_2$ is an artifact of the experiment. To collect the low temperature resistivity, an adiabatic refrigeration attachment was used. The process involves holding a 3 T magnetic field when cooling from 300 K. Once the machine is at 2 K, the field is linearly removed and the experiment begins recording. Since this sample has the lowest $T_c$, we suggest it also has a significantly low $H_c$ and the magnetic field lines still penetrate the sample at low fields and temperatures. The tail is an artifact of decreasing magnetic field and increasing temperature.

*Volume Dependence on Temperature*

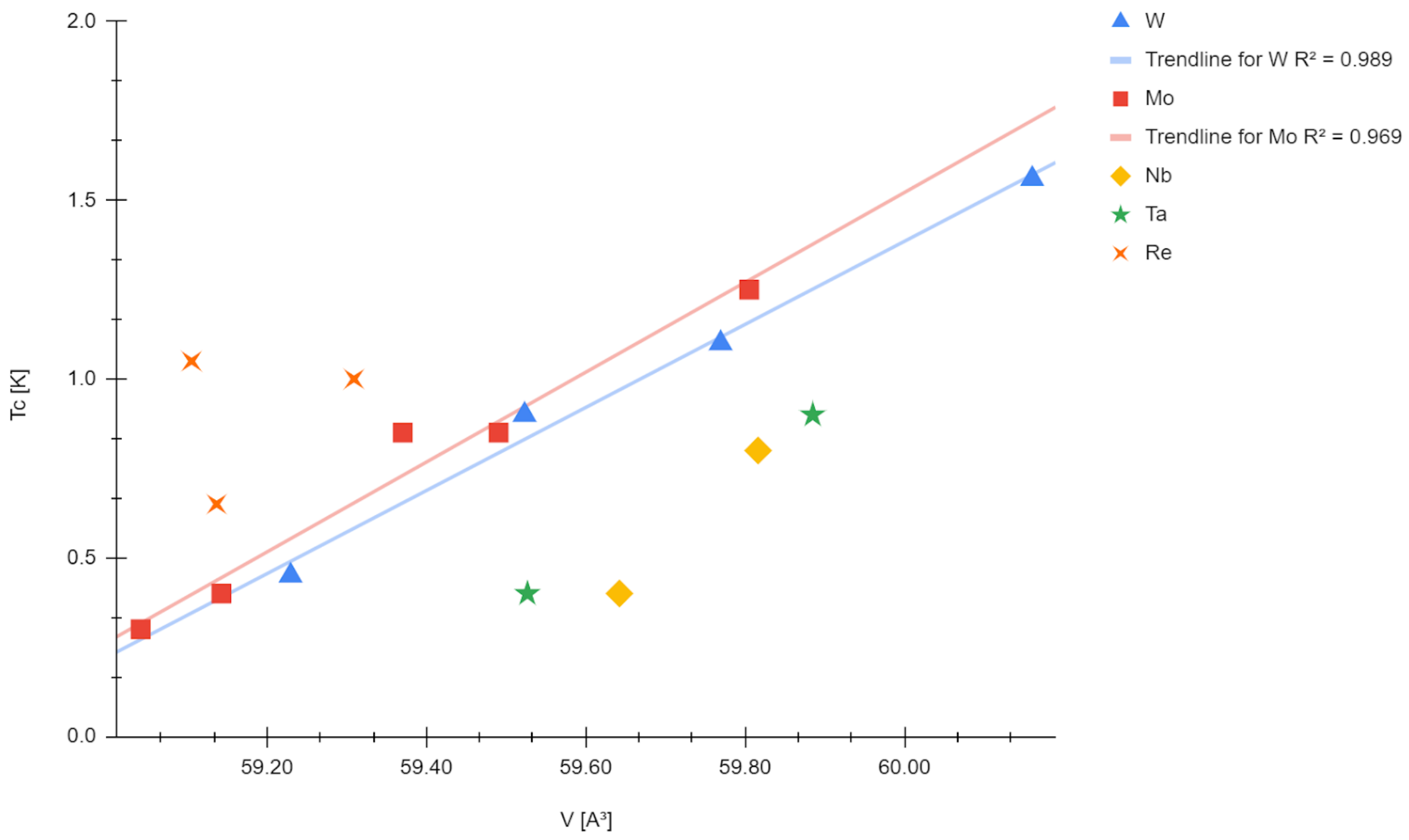

## List of values used in plots

| compound | structure | Tc | delta | delta_H | VEC | Source Title |
|---|---|---|---|---|---|---|
| Ta34Nb33Hf8Zr14Ti11 | bcc | 7.3 | 4.57 | 2.5 | 4.67 | Discovery of a superconducting high-entropy alloy |
| (TaNb)0.7(HfZrTi)0.3 | bcc | 8 | 4.35 | 2.38 | 4.7 | Effect of electron count and chemical complexity in the Ta-Nb-Hf-Zr-Ti high-entropy alloy superconductor. |
| (TaNb)0.67(HfZrTi)0.33 | bcc | 7.8 | 4.48 | 2.51 | 4.67 | Effect of electron count and chemical complexity in the Ta-Nb-Hf-Zr-Ti high-entropy alloy superconductor. |
| (TaNb)0.6(HfZrTi)0.4 | bcc | 7.6 | 4.72 | 2.72 | 4.6 | Effect of electron count and chemical complexity in the Ta-Nb-Hf-Zr-Ti high-entropy alloy superconductor. |
| (TaNb)0.5(HfZrTi)0.5 | bcc | 6.5 | 4.93 | 2.83 | 4.5 | Effect of electron count and chemical complexity in the Ta-Nb-Hf-Zr-Ti high-entropy alloy superconductor. |
| (TaNb)0.16(HfZrTi)0.84 | bcc | 4.5 | 4.65 | 1.52 | 4.16 | Effect of electron count and chemical complexity in the Ta-Nb-Hf-Zr-Ti high-entropy alloy superconductor. |
| (TaNbV)0.67(HfZrTi)0.33 | bcc | 4.3 | 6.41 | 0.49 | 4.67 | Effect of electron count and chemical complexity in the Ta-Nb-Hf-Zr-Ti high-entropy alloy superconductor. |
| (NbV)0.67(HfZrTi)0.33 | bcc | 7.2 | 7.05 | -0.15 | 4.67 | Effect of electron count and chemical complexity in the Ta-Nb-Hf-Zr-Ti high-entropy alloy superconductor. |
| (TaV)0.67(HfZrTi)0.33 | bcc | 4 | 7.05 | -0.6 | 4.67 | Effect of electron count and chemical complexity in the Ta-Nb-Hf-Zr-Ti high-entropy alloy superconductor. |
| (Sc0.33Cr0.67Nb)0.67(HfZrTi)0.33 | bcc | 5.6 | 9.15 | 0.2 | 4.67 | Isoelectronic substitutions and aluminium alloying in the Ta-Nb-Hf-Zr-Ti high-entropy alloy superconductor |
| (TaSc0.33Cr0.67)0.67(HfZrTi)0.33 | bcc | 4.4 | 9.15 | -0.54 | 4.67 | Isoelectronic substitutions and aluminium alloying in the Ta-Nb-Hf-Zr-Ti high-entropy alloy superconductor |
| (TaNb)0.67(Sc0.67Cr0.33ZrTi)0.33 | bcc | 7.5 | 5.69 | 4.25 | 4.67 | Isoelectronic substitutions and aluminium alloying in the Ta-Nb-Hf-Zr-Ti high-entropy alloy superconductor |
| (TaNb)0.67(HfSc0.67Cr0.33Ti)0.33 | bcc | 7.4 | 5.39 | 4.33 | 4.67 | Isoelectronic substitutions and aluminium alloying in the Ta-Nb-Hf-Zr-Ti high-entropy alloy superconductor |
| (TaNb)0.67(HfZrSc0.67Cr0.33)0.33 | bcc | 7.6 | 6.16 | 4.71 | 4.67 | Isoelectronic substitutions and aluminium alloying in the Ta-Nb-Hf-Zr-Ti high-entropy alloy superconductor |
| (Y0.33Mo0.67Nb)0.67(HfZrTi)0.33 | bcc | 4.7 | 8.96 | 6.82 | 4.67 | Isoelectronic substitutions and aluminium alloying in the Ta-Nb-Hf-Zr-Ti high-entropy alloy superconductor |
| (TaY0.33Mo0.67)0.67(HfZrTi)0.33 | bcc | 3.5 | 8.95 | 6.23 | 4.67 | Isoelectronic substitutions and aluminium alloying in the Ta-Nb-Hf-Zr-Ti high-entropy alloy superconductor |
| (TaNb)0.67(Y0.67Mo0.33ZrTi)0.33 | bcc | 7.6 | 7.27 | 7.44 | 4.67 | Isoelectronic substitutions and aluminium alloying in the Ta-Nb-Hf-Zr-Ti high-entropy alloy superconductor |
| (TaNb)0.67(HfY0.67Mo0.33Ti)0.33 | bcc | 6.7 | 7.08 | 7.54 | 4.67 | Isoelectronic substitutions and aluminium alloying in the Ta-Nb-Hf-Zr-Ti high-entropy alloy superconductor |
| (TaNb)0.67(HfZrY0.67Mo0.33)0.33 | bcc | 7.5 | 7.49 | 7.9 | 4.67 | Isoelectronic substitutions and aluminium alloying in the Ta-Nb-Hf-Zr-Ti high-entropy alloy superconductor |
| (Sc0.33Mo0.67Nb)0.67(HfZrTi)0.33 | bcc | 4.4 | 6.62 | 2.87 | 4.67 | Isoelectronic substitutions and aluminium alloying in the Ta-Nb-Hf-Zr-Ti high-entropy alloy superconductor |
| (TaSc0.33Mo0.67)0.67(HfZrTi)0.33 | bcc | 2.9 | 6.61 | 2.43 | 4.67 | Isoelectronic substitutions and aluminium alloying in the Ta-Nb-Hf-Zr-Ti high-entropy alloy superconductor |
| (TaNb)0.67(Sc0.67Mo0.33ZrTi)0.33 | bcc | 7.5 | 5.11 | 4.64 | 4.67 | Isoelectronic substitutions and aluminium alloying in the Ta-Nb-Hf-Zr-Ti high-entropy alloy superconductor |
| (TaNb)0.67(HfSc0.67Mo0.33Ti)0.33 | bcc | 6.6 | 4.79 | 4.71 | 4.67 | Isoelectronic substitutions and aluminium alloying in the Ta-Nb-Hf-Zr-Ti high-entropy alloy superconductor |
| (TaNb)0.67(HfZrSc0.67Mo0.33)0.33 | bcc | 7.5 | 5.59 | 5.14 | 4.67 | Isoelectronic substitutions and aluminium alloying in the Ta-Nb-Hf-Zr-Ti high-entropy alloy superconductor |
| Ta0.167Nb0.333Hf0.167Zr0.167Ti0.167 | bcc | 7.9 | 4.93 | 3 | 4.5 | Strongly correlated and strongly coupled s-wave superconductivity of the high entropy alloy Ta1/6Nb2/6Hf1/6Zr1/6Ti1/6 compound |
| Nb22.1Ta26.3Ti16.6Zr15.5Hf19.5 | bcc | 7.1 | 4.92 | 2.81 | 4.48 | Superconducting in equal molar NbTaTiZr-based high-entropy alloys |
| Nb21.5Ta18.1Ti15.9Zr14.4Hf16.6V13.5 | bcc | 5.1 | 6.22 | 1.25 | 4.53 | Superconducting in equal molar NbTaTiZr-based high-entropy alloys |
| NbTaTiZrFe | bcc | 6.9 | 8.06 | -10.08 | 5.2 | Superconducting in equal molar NbTaTiZr-based high-entropy alloys |
| NbReZrHfTi | bcc | 5.3 | 5.87 | -16.96 | 4.8 | Superconductivity in equimolar Nb-Re-Hf-Zr-Ti high entropy alloy |
| Hf21Nb25Ti15V15Zr24 | bcc | 5.3 | 6.8 | 0.94 | 4.4 | New high-entropy alloy superconductor Hf21Nb25Ti15V15Zr24 |
| Mo0.2375Re0.2375Ru0.2375Rh0.2375Ti0.05 | hcp | 3.6 | 2 | -13.79 | 7.32 | Superconductivity in high and medium entropy alloys based on MoReRu |
| Mo0.225Re0.225Ru0.225Rh0.225Ti0.1 | hcp | 4.7 | 2.55 | -18.25 | 7.15 | Superconductivity in high and medium entropy alloys based on MoReRu |
| Mo0.1Re0.1Ru0.55Rh0.1Ti0.15 | hcp | 2.1 | 3.16 | -22.97 | 7.2 | Superconductivity in high and medium entropy alloys based on MoReRu |
| Mo0.105Re0.105Ru0.527Rh0.105Ti0.158 | hcp | 2.2 | 3.21 | -23.72 | 7.16 | Superconductivity in high and medium entropy alloys based on MoReRu |
| Mo0.118Re0.118Ru0.47Rh0.118Ti0.176 | hcp | 2.5 | 3.31 | -25.23 | 7.06 | Superconductivity in high and medium entropy alloys based on MoReRu |
| Re0.56Nb0.11Ti0.11Zr0.11Hf0.11 | hcp | 4.4 | 6 | -28.1 | 5.79 | Superconductivity in a new hexagonal high-entropy alloy |
| Nb10Mo35Ru35Rh10Pd10 | hcp | 5.6 | 1.94 | -20.54 | 7.3 | Superconductivity in hexagonal Nb-Mo-Ru-Rh-Pd high-entropy alloys. |
| Nb15Mo32.5Ru32.5Rh10Pd10 | hcp | 6.2 | 2.2 | -23.93 | 7.2 | Superconductivity in hexagonal Nb-Mo-Ru-Rh-Pd high-entropy alloys. |
| Nb20Mo30Ru30Rh10Pd10 | hcp | 6.1 | 2.4 | -26.92 | 7.1 | Superconductivity in hexagonal Nb-Mo-Ru-Rh-Pd high-entropy alloys. |
| Nb5Mo35Re15Ru35Rh10 | hcp | 7.54 | 1.58 | -15.43 | 7.1 | Structural evolution and superconductivity tuned by valence electron concentration in the Nb-Mo-Re-Ru-Rh high-entropy alloys |
| Nb5Mo30Re20Ru35Rh10 | hcp | 6.69 | 1.6 | -14.61 | 7.15 | Structural evolution and superconductivity tuned by valence electron concentration in the Nb-Mo-Re-Ru-Rh high-entropy alloys |
| Nb5Mo25Re25Ru35Rh10 | hcp | 6.51 | 1.62 | -13.65 | 7.2 | Structural evolution and superconductivity tuned by valence electron concentration in the Nb-Mo-Re-Ru-Rh high-entropy alloys |
| Nb5Mo20Re30Ru35Rh10 | hcp | 5.46 | 1.64 | -12.55 | 7.25 | Structural evolution and superconductivity tuned by valence electron concentration in the Nb-Mo-Re-Ru-Rh high-entropy alloys |
| (ZrNb)0.2(MoReRu)0.8 | a-Mn | 4.2 | 5.41 | -24.55 | 6.5 | High-entropy alloy superconductors on an α-Mn lattice. |
| (ZrNb)0.1(MoReRu)0.9 | a-Mn | 5.3 | 4.11 | -18.26 | 6.75 | High-entropy alloy superconductors on an α-Mn lattice. |
| (HfTaWIr)0.4Re0.6 | a-Mn | 4 | 4.48 | -20.48 | 6.6 | High-entropy alloy superconductors on an α-Mn lattice. |
| (HfTaWIr)0.3Re0.7 | a-Mn | 4.5 | 3.96 | -16.09 | 6.7 | High-entropy alloy superconductors on an α-Mn lattice. |
| (HfTaWIr)0.25Re0.75 | a-Mn | 5.6 | 3.65 | -13.72 | 6.75 | High-entropy alloy superconductors on an α-Mn lattice. |
| (HfTaWPt)0.4Re0.6 | a-Mn | 4.4 | 4.37 | -22.32 | 6.7 | High-entropy alloy superconductors on an α-Mn lattice. |
| (HfTaWPt)0.3Re0.7 | a-Mn | 5.7 | 3.89 | -17.2 | 6.78 | High-entropy alloy superconductors on an α-Mn lattice. |
| Nb25Mo5Re35Ru25Rh10 | a-Mn | 4.66 | 2.48 | -25.85 | 6.9 | Structural evolution and superconductivity tuned by valence electron concentration in the Nb-Mo-Re-Ru-Rh high-entropy alloys |
| Nb25Mo10Re35Ru20Rh10 | a-Mn | 5.1 | 2.41 | -25.26 | 6.8 | Structural evolution and superconductivity tuned by valence electron concentration in the Nb-Mo-Re-Ru-Rh high-entropy alloys |
| Nb25Mo15Re35Ru15Rh10 | a-Mn | 5.1 | 2.34 | -24.39 | 6.7 | Structural evolution and superconductivity tuned by valence electron concentration in the Nb-Mo-Re-Ru-Rh high-entropy alloys |
| Ta5(Mo35W5)Re35Ru20 | sigma | 6.3 | 1.45 | -11.97 | 6.7 | Formation and superconductivity of single-phase high-entropy alloys with a tetragonal structure |
| Ta5(Mo30W10)Re35Ru20 | sigma | 6.2 | 1.45 | -11.62 | 6.7 | Formation and superconductivity of single-phase high-entropy alloys with a tetragonal structure |
| Ta5(Mo25W15)Re35Ru20 | sigma | 6.1 | 1.45 | -11.2 | 6.7 | Formation and superconductivity of single-phase high-entropy alloys with a tetragonal structure |
| Ta5(Mo20W20)Re35Ru20 | sigma | 5.7 | 1.45 | -10.92 | 6.7 | Formation and superconductivity of single-phase high-entropy alloys with a tetragonal structure |
| Ta5(Mo15W25)Re35Ru20 | sigma | 5.5 | 1.45 | -10.57 | 6.7 | Formation and superconductivity of single-phase high-entropy alloys with a tetragonal structure |
| Ta5(Mo10W30)Re35Ru20 | sigma | 5.5 | 1.45 | -10.22 | 6.7 | Formation and superconductivity of single-phase high-entropy alloys with a tetragonal structure |
| Ta5(Mo5W35)Re35Ru20 | sigma | 4.8 | 1.44 | -9.87 | 6.7 | Formation and superconductivity of single-phase high-entropy alloys with a tetragonal structure |
| (Ta7Mo33)W5Re35Ru20 | sigma | 6.1 | 1.59 | -12.99 | 6.68 | Formation and superconductivity of single-phase high-entropy alloys with a tetragonal structure |
| (Ta9Mo31)W5Re35Ru20 | sigma | 5.7 | 1.71 | -13.99 | 6.66 | Formation and superconductivity of single-phase high-entropy alloys with a tetragonal structure |
| (Ta11Mo29)W5Re35Ru20 | sigma | 5.3 | 1.82 | -14.97 | 6.64 | Formation and superconductivity of single-phase high-entropy alloys with a tetragonal structure |
| (Ta13Mo27)W5Re35Ru20 | sigma | 5.3 | 1.92 | -15.94 | 6.62 | Formation and superconductivity of single-phase high-entropy alloys with a tetragonal structure |
| Ta10Mo30Cr5Re35Ru20 | sigma | 4.79 | 2.62 | -14.56 | 6.65 | Superconductivity and paramagnetism in Cr-containing tetragonal high entropy alloys |
| Ta10Mo25Cr10Re35Ru20 | sigma | 4.41 | 3.21 | -14.31 | 6.65 | Superconductivity and paramagnetism in Cr-containing tetragonal high entropy alloys |
| Ta10Mo22Cr13Re35Ru20 | sigma | 3.98 | 3.49 | -14.16 | 6.65 | Superconductivity and paramagnetism in Cr-containing tetragonal high entropy alloys |
| W46.7Rh13.3Ir13.3Pt26.7 | fcc | 1.56 | 0.99 | -16.40 | 7.87 | This text |
| W42.9Rh14.3Ir14.3Pt28.6 | fcc | 1.10 | 1.02 | -16.17 | 8.00 | This text |
| W37.9Mo5Rh14.3Ir14.3Pt28.6 | fcc | 1.25 | 1.03 | -16.94 | 8.00 | This text |
| W38.5Rh15.4Ir15.4Pt30.8 | fcc | 0.90 | 1.06 | -15.68 | 8.15 | This text |
| W33.5Mo5Rh15.4Ir15.4Pt30.8 | fcc | 0.85 | 1.07 | -16.51 | 8.15 | This text |
| W28.5Mo10Rh15.4Ir15.4Pt30.8 | fcc | 0.85 | 1.07 | -17.34 | 8.15 | This text |
| W33.3Rh16.7Ir16.7Pt33.3 | fcc | 0.45 | 1.11 | -14.77 | 8.33 | This text |
| W28.3Mo5Rh16.7Ir16.7Pt33.3 | fcc | 0.40 | 1.11 | -15.67 | 8.33 | This text |
| W23.3Mo10Rh16.7Ir16.7Pt33.3 | fcc | 0.30 | 1.11 | -16.57 | 8.33 | This text |
| W33.5Ta5Rh15.4Ir15.4Pt30.8 | fcc | 0.9 | 1.45 | -21.19 | 8.11 | This text |
| W28.3Ta5Rh16.7Ir16.7Pt33.3 | fcc | 0.4 | 1.48 | -20.63 | 8.28 | This text |
| W33.5Nb5Rh15.4Ir15.4Pt30.8 | fcc | 0.8 | 1.44 | -21.38 | 8.11 | This text |
| W28.3Nb5Rh16.7Ir16.7Pt33.3 | fcc | 0.4 | 1.46 | -20.82 | 8.28 | This text |
| W33.5Re5Rh15.4Ir15.4Pt30.8 | fcc | 1 | 1.07 | -14.26 | 8.21 | This text |
| W28.5Re10Rh15.4Ir15.4Pt30.8 | fcc | 1.05 | 1.08 | -12.75 | 8.26 | This text |
| W28.3Re5Rh16.7Ir16.7Pt33.3 | fcc | 0.65 | 1.11 | -13.16 | 8.38 | This text |